# A van der Waals antiferromagnetic topological insulator with weak interlayer magnetic coupling

Chaowei Hu[1], Kyle N. Gordon[2], Pengfei Liu[3], Jinyu Liu[1], Xiaoqing Zhou[2], Peipei Hao[2], Dushyant Narayan[2], Eve Emmanouilidou[1], Hongyi Sun[3], Yuntian Liu[3], Harlan Brawer[1], Arthur P. Ramirez[4], Lei Ding[5], Huibo Cao[5], Qihang Liu[3,6‡], Dan Dessau[2,7†], and Ni Ni[1*]

[1]Department of Physics and Astronomy and California NanoSystems Institute, University of California, Los Angeles, CA 90095, USA

[2]Department of Physics, University of Colorado, Boulder, CO 80309, USA

[3]Shenzhen Institute for Quantum Science and Technology and Department of Physics, Southern University of Science and Technology, Shenzhen, 518055, China

[4]Department of Physics, University of California, Santa Cruz, CA 95064, USA

[5]Neutron Scattering Division, Oak Ridge National Laboratory, Oak Ridge, Tennessee 37831, USA

[6]Guangdong Provincial Key Laboratory for Computational Science and Material Design, Southern University of Science and Technology, Shenzhen 518055, China

[7]Center for Experiments on Quantum Materials, University of Colorado, Boulder, CO 80309, USA

* nini@physics.ucla.edu

† dessau@colorado.edu

‡ liuqh@sustech.edu.cn

# ABSTRACT

Magnetic topological insulators (TI) provide an important material platform to explore quantum phenomena such as quantized anomalous Hall (QAH) effect and Majorana modes, etc. Their successful material realization is thus essential for our fundamental understanding and potential technical revolutions. By realizing a bulk van der Waals material MnBi$_4$Te$_7$ with alternating septuple [MnBi$_2$Te$_4$] and quintuple [Bi$_2$Te$_3$] layers, we show that it is ferromagnetic in plane but antiferromagnetic along the $c$ axis with an out-of-plane saturation field of ~ 0.22 T at 2 K. Our angle-resolved photoemission spectroscopy measurements and first-principles calculations further demonstrate that MnBi$_4$Te$_7$ is a Z$_2$ antiferromagnetic TI with two types of surface states associated with the [MnBi$_2$Te$_4$] or [Bi$_2$Te$_3$] termination, respectively. Additionally, its superlattice nature may make various heterostructures of [MnBi$_2$Te$_4$] and [Bi$_2$Te$_3$] layers possible by exfoliation. Therefore, the low saturation field and the superlattice nature of MnBi$_4$Te$_7$ make it an ideal system to investigate rich emergent phenomena.

**Introduction**

Magnetic topological insulators (MTIs), including Chern insulators with a Z-invariant and antiferromagnetic (AFM) topological insulators (TIs) with a $Z_2$-invariant, provide fertile ground for the exploration of emergent quantum phenomena such as the quantum anomalous Hall (QAH) effect, Majorana modes, the topological magnetoelectric effect, the proximity effect, etc[1,2]. In the two-dimensional (2D) limit of ferromagnetic (FM) TIs, the QAH effect arising from chiral edge states exists under zero external magnetic fields, which has been experimentally observed in doped FM TI $Cr_{0.15}(Bi_{0.1}Sb_{0.9})_{1.85}Te_3$ thin films[3]. However, the unavoidable sample inhomogeneity in doped materials restrains the investigation of associated emergent phenomena below temperatures of hundreds of mK[2]. Stoichiometric MTIs are expected to have homogeneous electronic and magnetic properties, which may provide new opportunities to study the QAH effect. Recently, $MnBi_2Te_4$ was discovered to be an intrinsic AFM TI[4–28]. In its 2D limit, quantized Hall conductance originating from the topological protected dissipationless chiral edge states was realized in few-layer slabs[15,16]. However, probably because the uncompensated AFM spin configuration cannot provide enough Zeeman field to realize the band inversion in only one spin channel, to observe such a QAH effect, a high magnetic field of 12 T at 4.5 K or 6 T at 1.5 K is required to fully polarize the AFM spins into a forced FM state[15,16].

A FM state is crucial to realize the QAH effect experimentally[15]; however, as we await an ideal candidate that has both TI and FM properties, an intrinsic AFM TI with low saturation fields and clean band structure where only non-trivial bands cross the Fermi level can also provide a good material platform. By this, the QAH effect may be realized with higher temperatures and reasonably low magnetic fields, which allows us to study their associated emergent phenomena at more accessible conditions. How can we realize such intrinsic AFM TIs? Recall that $MnBi_2Te_4$ crystalizing in the $GeBi_2Te_4$ structure with septuple layers (SL) of $[MnBi_2Te_4]$ is an AFM material with in-plane FM and out-of-plane AFM exchange interaction. Hence, based on the SL building block, one strategy to achieve AFM with small saturation fields or even FM is to reduce the interlayer Mn-Mn exchange interaction by increasing the interlayer distance with extra spacer layers added. Structurally, SL blocks have great compatibility with quintuple-layered (QL) blocks of $[Bi_2Te_3]$, whose bulk form is a time-reversal-preserved TI[29]. As an example, GeBi$_4$Te$_7$ with alternating [GeBi$_2$Te$_4$] and [Bi$_2$Te$_3$] building blocks has been synthesized[29]. This superior compatibility provides us with flexible structural control to achieve our goal. Furthermore, not only can such superlattices manifest weak interlayer magnetic coupling, but they can also serve as natural heterostructures by exfoliation, which may enable the realization of various topological states.

The exploration of the MnTe-Bi$_2$Te$_3$ ternary system[30] has shown that MnBi$_{2n}$Te$_{3n+1}$ ($n$ = 1, 2 and 3) series exist with alternating [MnBi$_2$Te$_4$] and ($n$-1)[Bi$_2$Te$_3$] layers. In this work, we focus on MnBi$_4$Te$_7$ ($n$ = 2) with a hexagonal superlattice crystal structure of alternate stacking of one [MnBi$_2$Te$_4$] SL and one [Bi$_2$Te$_3$] QL, as shown in Fig. 1(a). Through our transport, thermodynamic, angle-resolved photoemission spectroscopy (ARPES) and density functional theory (DFT) calculations, we discovered that MnBi$_4$Te$_7$ is a Z$_2$ AFM TI with an out-of-plane saturation field as low as 0.22 T at 2 K, 40 times lower than that of MnBi$_2$Te$_4$. Furthermore, the natural-heterostructure-like construction of MnBi$_4$Te$_7$ can host two distinct (001) surface states. For the [Bi$_2$Te$_3$] termination, clean gapped surface states are observed as has long been desired; while for the [MnBi$_2$Te$_4$] termination, nearly gapless surface Dirac cone is observed, similar to the case of the MnBi$_2$Te$_4$ compound [20–23]. Our finding provides a superior new material realization to explore the QAH effect, quantum spin Hall (QSH) effect and associated phenomena[31].

**Results**

**A-type antiferromagnetism in MnBi$_4$Te$_7$ with strong FM fluctuations and weak interlayer exchange interaction.** Figure 1(b) shows the (0 0 $l$) x-ray diffraction peaks of a representative crystal, which can be well indexed by the MnBi$_4$Te$_7$ crystal structure[30]. The Rietveld refinement of the powder X-ray diffraction pattern agrees well the MnBi$_4$Te$_7$ structure model[30] and suggests Bi$_2$Te$_3$ is the only impurity inside with a molar ratio of 14% (See Fig. S1). The refined lattice parameters are $a$ = 4.3454(5) Å, and $c$ = 23.706(4) Å, indicating the distance between two adjacent Mn layers in MnBi$_4$Te$_7$ is 23.706(4) Å, much longer than the 13.8 Å of MnBi$_2$Te$_4$. The inset of Fig. 1(b) shows a picture of a MnBi$_4$Te$_7$ single crystal against a 1-mm scale, where the shiny cleaved $ab$ surface can be seen.

The magnetic properties are depicted in Figs. 1(c)-(e). Figure 1(c) presents the field-cooled (FC) magnetic susceptibility data of $\chi^{ab}$ (H ∥ ab) and $\chi^c$ (H ∥ c) measured at 0.1 T. The abrupt halt in the rise of $\chi^c$ on cooling suggests the onset of AFM ordering, similar to that seen in other vdW antiferromagnets $MnBi_2Te_4$ and $CrCl_3$,[9,18,21] but different from the FM one[22], suggesting that long range AFM ordering takes place at 13 K. This is consistent with the specific heat measurement as shown in the inset of Fig. 1(c), where a specific heat anomaly associated with the AFM transition emerges at 13 K. As seen from Fig. 1(c), fitting the inverse susceptibilities up to 80 K to the Curie-Weiss law results in Weiss temperatures of $\theta_w^{ab}$ = 11.5 K, $\theta_w^c$ = 12.2 K, $\theta_w^{ave}$ = 11.7 K, and effective moments of $\mu_{eff}^{ab}$ = 5.4$\mu_B$/Mn, $\mu_{eff}^c$ = 5.1$\mu_B$/Mn and $\mu_{eff}^{ave}$ = 5.3$\mu_B$/Mn. These values indicate magnetic isotropy above $T_N$ and thus negligible single ion anisotropy in the material. Despite the fact that $MnBi_4Te_7$ is AFM below 13 K, the positive $\theta_w^{ave}$ of 11.7 K suggests strong ferromagnetic (FM) exchange interactions. Recall that $MnBi_2Te_4$ has a much higher $T_N$ of 25 K and a much lower $\theta_w$ of 3~6 K[9,18], this may indicate that the energy scales of the FM and AFM exchange interaction are much closer in $MnBi_4Te_7$. This is consistent with the fact that the extra insulating [$Bi_2Te_3$] layer reduces the interlayer exchange interaction between adjacent Mn layers as we initially designed. The AFM orders of both $MnBi_2Te_4$ and $MnBi_4Te_7$ are formed under the superexchange scenario, where the magnetic interaction between the adjacent Mn layers is mediated by the electrons of the common neighbors. Despite the long distance between the adjacent Mn layers (23.706(4) Å), our DFT calculation reveals an A-type AFM configuration in $MnBi_4Te_7$ with the interlayer exchange coupling about -0.15 meV/Mn, which is about one order of magnitude smaller than the counterpart of $MnBi_2Te_4$. More details are given in Supplementary Note II.

Figures 1(d)-(e) present the hysteresis loops of isothermal magnetization data for $M^c(H)$ (H ∥ c) and $M^{ab}(H)$ (H ∥ ab), respectively. As shown in Fig. 1(d), in sharp contrast to $MnBi_2Te_4$ where a spin-flop transition takes place at 3.5 T and saturates at 8 T in $M^c(H)$[9,17], $MnBi_4Te_7$ undergoes a first-order spin-flip transition with hysteresis starting at a much lower field of $H_f$ = 0.15 T. It quickly enters the forced FM state and saturates at $H_c$ = 0.22 T. The small saturation field again indicates weaker interlayer AFM exchange interactions than in MnBi$_2$Te$_4$. Upon warming up to 10 K, the hysteresis area is gradually reduced to zero, but $H_c$ remains little changed, indicating a sharp triggering of the spin-flipping between 10 K and T$_N$. With H $\parallel ab$, the saturation field is 1.0 T, indicating the $c$ axis as the magnetic easy axis and likely Ising form. As shown in Fig. 1(e), the saturation moment is 3.5$\mu_B$/Mn at 7 T, which is very similar to the value of 3.6$\mu_B$/Mn[9,18] in MnBi$_2$Te$_4$ but smaller than the DFT calculated value of 4.6 $\mu_B$/Mn. The reduced Mn saturation moments in this family may arise from Mn disorders, as suggested in MnBi$_2$Te$_4$[10].

Figure 1(f) shows the temperature dependent in-plane ($\rho_{xx}$) and out-of-plane resistivity ($\rho_{zz}$). Above 20 K, both $\rho_{xx}$ and $\rho_{zz}$ decrease nearly linearly upon cooling with $\rho_{zz}/\rho_{xx}$ ~ 53 at 300 K (see Fig. S2), suggesting a large transport anisotropy that is consistent with its vdW nature. With further cooling, $\rho_{xx}$ and $\rho_{zz}$ increase slightly, which is likely caused by the enhanced scattering from spin fluctuations, a phenomenon frequently observed in low dimensional magnetic materials[34,35]. Then at 13 K, a sudden drop of $\rho_{xx}$ and a sharp increase of $\rho_{zz}$ are observed. This is in agreement with the A-type magnetic structure shown in Fig. 1(a) since the antiparallel alignment of Mn moments can reduce the conductivity via spin-slip scattering, while parallel alignment of the Mn moments will eliminate such scattering and thus enhance the conductivity [34].

Figure 1 (g) shows the transverse magnetoresistance (TMR), defined as $MR = (\rho_{xx}(H) - \rho_{xx}(0))/ \rho_{xx}(0)$. The main feature of the figure is the overall "W" shape of the TMR. The "W" shape becomes deeper with increasing temperature, with the largest negative TMR of 8% appearing at 12 K, which is close to T$_N$. Above T$_N$, it starts to become shallower and finally transforms into an ordinary parabolic shape at 50 K. The overall "W" shape can be understood in the framework of FM fluctuations. Above 50 K, the lack of magnetic fluctuations leads to the parabolic TMR. Upon cooling, FM fluctuations begin to appear and become increasingly stronger with maxima around T$_N$. As a result, the summation of the positive parabolic TMR and the negative TMR arising from the FM fluctuations under fields leads to a progressively deeper "W" shape of TMR upon cooling. Below T$_N$, the FM fluctuations are reduced, but still with a strong presence, leading to the shallower "W" shape under field.

The spin-flip transition strongly affects the transport properties, as shown in Fig. 2. $\rho_{xx}(H)$, $\rho_{zz}(H)$ and $\rho_{xy}(H)$ follow the same hysteresis as that in $M(H)$. With $H \parallel c$, the transverse magnetoresistivity of $\rho_{xx}$ with $I \parallel ab$ (Fig. 2(a)) and the longitudinal magnetoresistivity of $\rho_{zz}$ with $I \parallel c$ (Fig. 2(b)) slightly change between 0 T to $H_f$. Then up to $H_c$, since the system enters the forced FM state and the loss of spin scattering occurs, $\rho_{xx}$ drops by 3.8% whereas $\rho_{zz}$ decreases by 34%. With $H \parallel ab$, up to the saturation field of 1.0 T, $\rho_{zz}$ (Fig. 2(e)) decreases by 39% whereas $\rho_{xx}$ (Fig. 2(f)) drops by 2.6%. Our data show that the transition from AFM to FM spin alignment along the $c$ axis has much stronger effect on $\rho_{zz}$ than $\rho_{xx}$. MnBi$_4$Te$_7$ displays evident anomalous Hall effect (AHE) as seen in the bottom panel of Fig. 2(a). Our $\rho_{xy}(H)$ is linear up to 9 T above 50 K (see Fig. S3) suggesting single band transport here. Using $n = H/e\rho_{xy}$, our 50 K data (see Fig. S3) corresponds to an electron carrier density of $2.84 \times 10^{20}$ cm$^{-3}$, similar to that of MnBi$_2$Te$_4$[17, 18, 23]. Our Hall resistivity below 13 K can be described by $\rho_{xy} = R_0 H + \rho_{xy}^A$, where the $R_0 H$ is the trivial linear contribution and $\rho_{xy}^A$ represents the anomalous Hall resistivity. At 2 K, $\rho_{xy}^A$ is extracted to be 3.3 μΩ cm, which is half of the one in MnBi$_2$Te$_4$[17]. Consequently, the anomalous Hall conductivity $\sigma_{xy}^A$ ($= \rho_{xy}^A/\rho_{xx}^2$) is 25.5 Ω$^{-1}$ cm$^{-1}$ and the anomalous Hall angle (AHA ~ $\rho_{xy}^A/\rho_{xx}$) is ~ 1%.

**Z$_2$ AFM TI predicted by theoretical calculation**. MnBi$_4$Te$_7$ crystalizes in the space group (*G*) *P-3m1* (No. 164). By taking into account the A-type AFM, the primitive cell doubles along the $c$ axis, rendering a magnetic space group $P_c$-3c1 (No. 165.96) under the Belov-Neronova-Smirnova notation[37], as shown in Fig. 1(a). This magnetic space group is derived from its nonmagnetic space group by adding an extra sublattice generated by an operation that combines time-reversal *T* with a fractional translation $\tau_{1/2}$. Then the full magnetic group is built as $G_M = G + GS$, where *S* is a combinatory symmetry $S = T\tau_{1/2}$ with $\tau_{1/2}$ the half translation along the $c$ axis of the AFM primitive cell. Although the explicit *T*- symmetry is broken, the *S* symmetry (also referred to nonsymmorphic time-reversal[38]) still exists in bulk MnBi$_4$Te$_7$. In addition, MnBi$_4$Te$_7$ has inversion symmetry *P*, while the square of the symmetry operator *PS* equals -1 at an arbitrary *k* in momentum space. Therefore, analogous to TI with *T*-symmetry where Kramer's degeneracy is induced by $T^2$ = -1, in MnBi$_4$Te$_7$ the existence of the *PS* symmetry ensures an equivalent Kramer's degeneracy in the whole Brillion zone, and thus a $Z_2$ topological classification.

Figure 3(a) shows the calculated band structure of bulk AFM MnBi$_4$Te$_7$ with the presence of spin-orbit coupling. The conduction band minimum is located at the Γ point, while the valence band maximum in the vicinity of Γ shows a slightly curved feature. The calculated bulk band gap is about 160 meV. The projection of band eigenstates onto the *p*-orbitals of Bi and Te (as indicated by the green and red coloring) clearly indicates an inverted order between several conduction and valence bands around the Γ point, which is strong evidence of the possible nontrivial topological nature. On the other hand, the Mn-$3d^5$ states form nearly flat bands far away from the Fermi level (see Fig. S5), indicating that the main effect of Mn is to break *T*-symmetry by introducing staggered Zeeman field into the low-energy Hamiltonian.

To determine the topological properties of AFM MnBi$_4$Te$_7$, we first apply the Fu-Kane formula[39] to calculate the $Z_2$ invariant. The topological insulator phase of antiferromagnetic materials is protected by *S* symmetry, under which there are only four invariant *k*-points forming a 2D plane in the momentum space. Thus, analogous to weak $Z_2$ indices in nonmagnetic materials, the *S* symmetry indeed protects weak $Z_2$ topological phases in antiferromagnetic materials. In AFM MnBi$_4$Te$_7$, four TRIM points, including Γ(0, 0, 0) and three equivalent *M* (π, 0, 0), need to be considered here with $\boldsymbol{k} \cdot \tau_{1/2} = n\pi$. Due to the abovementioned band inversion at the Γ point, we find that the parities for the occupied bands at Γ are opposite to that of the other three *M* points, indicating a nontrivial $Z_2$ = 1. To verify our results, we also calculate the evolution of Wannier charge centers (WCCs) using the Wilson loop approach[40]. As show in Fig. 3(b), the largest gap function and the WCCs line cross each other an odd number of times through the evolution, confirming that MnBi$_4$Te$_7$ is indeed a $Z_2$ AFM topological insulator. Compared with TIs with *T*-symmetry, the protection of gapless surface states in AFM TIs requires that the cleaved surface respects $S$ symmetry that contains translation along the $c$ axis. Figure 3(c) clearly shows the gapless surface Dirac cone at the $\Gamma$ point for the (010) surface, partially validating the bulk-surface correspondence of MnBi$_4$Te$_7$ as an AFM TI. The easy-cleaved (001) plane, where the $S$ symmetry is broken, are measured by ARPES and compared with our theoretical calculations, as discussed in the following.

**Surface and bulk states measured by ARPES.** In contrast to the recently discovered AFM TI MnBi$_2$Te$_4$ where only one type of surface termination exists, MnBi$_4$Te$_7$ can terminate on two different sub-lattice surfaces on the (001) plane, i.e., the [Bi$_2$Te$_3$] QL termination and the [MnBi$_2$Te$_4$] SL termination, resulting in different surface states. ARPES with 47 eV, linear horizontal polarized light and a small beam spot reveals two different types of *E-k* maps by scanning across different parts of the sample in real space, as plotted in Figs. 4(d,e) and Figs. 4(h,i). There are several distinguishing features between the two types of surface spectra: Figs. 4(h,i) appear to show a gap with massive quasiparticles while Figs. 4(d,e) show a sharp Dirac-like crossing, possibly with a small gap. The spectra of Figs. 4(d,e) are reminiscent of recent high resolution ARPES spectra of the MnBi$_2$Te$_4$ compound[20–23] that show Dirac-like spectra, and we assign these states to the [MnBi$_2$Te$_4$] SL termination, while we assign the other set of surface states to the [Bi$_2$Te$_3$] QL termination.

On these two terminations, symmetry operations combined with $\tau_{1/2}$ are not preserved. In the ideal case that the surface magnetic structure perfectly inherits the bulk property, due to the A-type out-of-plane magnetization of the Mn sublayers, the gapped surface states are described by adding an exchange term to the ordinary Rashba-type surface Hamiltonian for TI with $T$ symmetry, i.e., $H_{surf}(\boldsymbol{k}) = (\sigma_x k_y - \sigma_y k_x) + m_{S/Q}\sigma_z$, where $\sigma$ is the Pauli matrix for spin, and $m_{S/Q}$ the surface exchange field that distinguishes the [MnBi$_2$Te$_4$] SL and [Bi$_2$Te$_3$] QL surfaces. Our calculation shows that the surface state terminated at the [Bi$_2$Te$_3$] QL has a massive Dirac cone with a surface gap around 60 meV (see Fig. 4(f,g)), and an overall structure that agrees very well with the experimental data of Fig. 4(h,i), confirming the assignment of the experimental data as arising from the [Bi$_2$Te$_3$] QL termination. When comparing Fig. 4(i) with the bulk states calculated by DFT (Fig. 4 (b)), we can easily distinguish the surface states from the bulk states. To measure gap sizes in Fig. 4(i), we extract an energy distribution curve (EDC) at the Γ point and fit it to several Voigt profiles, as shown in Fig. 4(j). We find that despite the appearance of some spectral weight in the gapped region in Fig. 4(i), the EDC does not show any signature of a peak in the gapped region, indicating that the surface state is gapped by approximately 100 meV while the bulk gap is nearly 225 meV.

The equivalent calculation on the [MnBi$_2$Te$_4$] SL termination is shown in Figs. 4(a,c) and does not agree well with the experimental data of Fig. 4(d,e). While the theory shows that surface states merge with the bulk valence bands, the experiment suggests a Dirac-like structure inside the gap. By taking full account of experimental resolution functions in both momentum directions and in energy, the ARPES data are consistent with either no gap or a maximum gap size of 10 meV. More details are given in Supplementary Note III. A similar feature, i.e., nearly gapless surface Dirac cone at the SL termination, was observed recently in MnBi$_2$Te$_4$ single crystals[20–23], where the deviation between ARPES and DFT calculation is suggested to be due to the surface-mediated spin reconstruction at the top layers of the [MnBi$_2$Te$_4$] SL termination.

Figure 5(a,b) show stacks of measured isoenergy surfaces for the [MnBi$_2$Te$_4$] SL and [Bi$_2$Te$_3$] QL terminations over a wide range of energies both above and below the Dirac point, while Fig. 5c shows equivalent calculations for the [Bi$_2$Te$_3$] QL termination. The six-fold symmetric isoenergy surfaces are seen in all cases, including the hexagonal warping or snow-flake effect[41]. We comment that while both terminations collapse to a single resolution-limited point in *k*-space in the middle panels in Figs. 5 (a, b), this is expected whether or not there is a gapless or gapped Dirac point, due to the broad energy band width of the nearby valence and conduction bands (see Fig. 4j).

**Discussion**

The vdW AFM TI MnBi$_4$Te$_7$ single crystal reported here is in fact a 1:1 superlattice composing the building blocks of AFM TI [MnBi$_2$Te$_4$] and *T*-invariant TI [Bi$_2$Te$_3$]. Our realization of the superlattice design has three advantages. First, as discussed above, it serves as a "buffer layer" that separates and thus effectively decreases the AFM coupling between the two neighboring [MnBi$_2$Te$_4$] SLs, leading to a weaker magnetic field to trigger the QAH. Second, by interlayer coupling between [Bi$_2$Te$_3$] QL and the adjacent [MnBi$_2$Te$_4$] SLs, the SOC-induced nontrivial topology of [Bi$_2$Te$_3$] ensures the band inversion in the 2D limit. As a result, QAH is well expected in few-layer MnBi$_4$Te$_7$. Third, when MnBi$_4$Te$_7$ is exfoliated into the 2D limit, natural heterostructures are made, which provides more 2D configurations than MnBi$_2$Te$_4$ or Bi$_2$Te$_3$ single crystal since the latter ones are only stacked by one type of building block. One can exfoliate MnBi$_4$Te$_7$ with designed termination, with different film thickness and magnetization (require a low magnetic field). For example, two types of three-layer systems with distinct topological properties, [MnBi$_2$Te$_4$]/[Bi$_2$Te$_3$]/[MnBi$_2$Te$_4$] and [Bi$_2$Te$_3$]/[MnBi$_2$Te$_4$]/[Bi$_2$Te$_3$], should be easily obtained by exfoliation. Recent calculations[31] show that [MnBi$_2$Te$_4$]/[Bi$_2$Te$_3$]/[MnBi$_2$Te$_4$] is a QAH insulator if a small magnetic field around 0.2 T is applied to stabilize the forced FM phase. On the other hand, [Bi$_2$Te$_3$]/[MnBi$_2$Te$_4$]/[Bi$_2$Te$_3$] is suggested to be a QSH insulator with time-reversal breaking[33] which cannot be achieved from the thin films of either MnBi$_2$Te$_4$ or Bi$_2$Te$_3$. Therefore, the 2D version exfoliated from bulk vdW TI MnBi$_4$Te$_7$ paves an avenue to chase the long-sought emergent properties such as QAH effect and QSH effect. As the foundation of engineering 2D heterostructures, such topological vdW materials could open up unprecedented opportunities in discovering novel fundamental physics as well as making new quantum devices[42].

**Methods**

**Sample growth and characterization.** Single crystals of MnBi$_4$Te$_7$ were grown using self-flux[11]. Mn, Bi and Te elements are mixed so the molar ratio of MnTe: Bi$_2$Te$_3$ is 15:85. The mixtures are loaded in a 2mL crucibles, sealed in quartz tube, heated to and held at 900°C for 5 hours. After a quick cooling to 595°C, the mixtures are slowly cooled down to 582°C over one to three days, where sizable single crystals are obtained after centrifuging. Although Bi$_2$Te$_3$ is the inevitable side product, we can differentiate MnBi$_4$Te$_7$ pieces by measuring their (0 0 $l$) diffraction peaks. In each growth, a few sizable plate-like MnBi$_4$Te$_7$ single crystals with typical dimensions of $3 \times 3 \times 0.5\ mm^3$ were obtained. To confirm the phase, X-ray diffraction data were collected using a PANalytical Empyrean diffractometer (Cu Kα radiation). Samples used for powder X-ray diffraction were ground into powder inside acetone to reduce the preferred orientation. Electric resistivity and heat capacity data were measured in a Quantum Design (QD) DynaCool Physical Properties Measurement System (DynaCool PPMS). The magnetization data were measured in a QD Magnetic Properties Measurement System (QD MPMS). All magnetic data were calculated assuming the molar ratio between $MnBi_4Te_7$ and $Bi_2Te_3$ impurity is 86:14 in the sample suggested by powder X-ray refinement (Fig. S1). Magnetic data measured for H || c were corrected with a demagnetization factor.

**ARPES measurements.** ARPES measurements on single crystals of $MnBi_4Te_7$ were carried out at the Advanced Light Source beamline 7.0.2 with photon energies between 40 and 55 eV with linear horizontal polarization. Single crystal samples were top-posted on the (001) surface, and cleaved in-situ in an ultra-high vacuum better than $4 \times 10^{-11}$ Torr and a temperature of 15 K. ARPES spectra were taken at 12 K, slightly smaller than 13 K, the Neel temperature. As the cleaved terrain is expected to consist of patches of exposed $[Bi_2Te_3]$ QL and $[MnBi_2Te_4]$ SL, to eliminate the effect of possible QL and SL mixing on the ARPES data, we scanned a 1 mm square surface of the sample in 50 um steps with a 50 um beam spot and collected spectra from over 200 different spots on the sample. We looked at each spectrum, finding many regions with clear, sharp features. We also narrowed the beam spot down to 20 x 20 um and scanned more finely, in 15 um steps, in smaller regions of interest. We found that there were regions on the order of 50 x 50 um that were spectroscopically stable, meaning the ARPES spectra were not changing from spot to spot. We then took our data with a 20 x 20 um beam spot and studied the centroid of the spectroscopically stable regions, which we believe will minimize any contamination due to another surface.

**First-principles calculations.** We apply density functional theory (DFT) by using the projector-augmented wave (PAW) pseudopotentials[43] with the exchange-correlation of Perdew-Burke-Ernzerhof (PBE) form[44] and GGA+U[45] approach within the Dudarev scheme[37] as implemented in the Vienna ab-initio Simulation Package (VASP)[38]. The energy cutoff is chosen 1.5 times as large as the values recommended in relevant pseudopotentials.

The U value is set to be 5 eV[6]. The $k$-points-resolved value of BZ sampling is $0.02 \times 2\pi/\text{Å}$. The total energy minimization is performed with a tolerance of $10^{-6}$ eV. The crystal structure and atomic position are fully relaxed until the atomic force on each atom is less than $10^{-2}$ eV Å. SOC is included self-consistently throughout the calculations. We constructed Wannier representations[46,47] by projecting the Bloch states from the DFT calculations of bulk materials onto the Mn-3$d$, Bi-6$p$ and Te-5$p$ orbitals. The band spectra of the surface states are calculated in the tight-binding models constructed by these Wannier representations and by the iterative Green's function technique as implemented in WannierTools package[48].

**Acknowledgments**

We thank Paul C. Canfield, Dr. Quansheng Wu, Suyang Xu, Filip Ronning and Chris Regan for helpful discussions, and Chris Jozwiak and Roland Koch at the Advanced Light Source for experimental help. Work at UCLA and UCSC were supported by the U.S. Department of Energy (DOE), Office of Science, Office of Basic Energy Sciences under Award Number DE-SC0011978 and DE-SC0017862, respectively. Work at SUSTech was supported by the NSFC under Grant No. 11874195, the Guangdong Provincial Key Laboratory of Computational Science and Material Design under Grant No. 2019B030301001 and Center for Computational Science and Engineering of Southern University of Science and Technology. HC acknowledges the support from US DOE BES Early Career Award KC0402010 under Contract DE-AC05- 00OR22725. This research used resources of the Advanced Light Source, which is a DOE Office of Science User Facility under contract no. DE-AC02-05CH11231.

**Author contributions**

N. N. conceived the idea and organized the research. N. N., Q. L. and D. D. supervised the research. C. H., J. L., E. E., H. B. and N. N. grew the bulk single crystal and carried out X-ray and transport measurements. K. G., X. Z., P. H., D. N. and D. D. carried out the ARPES measurements and data analysis. Q. L., P. L., H. S. and Y. L. performed the first-principles calculations. A. R. and C. H performed magnetic measurements. H. C., L. D. and C. H. carried out structure determination. N. N., Q. L., D. D and K. G. prepared the manuscript with contributions from all authors.

Figure 1.

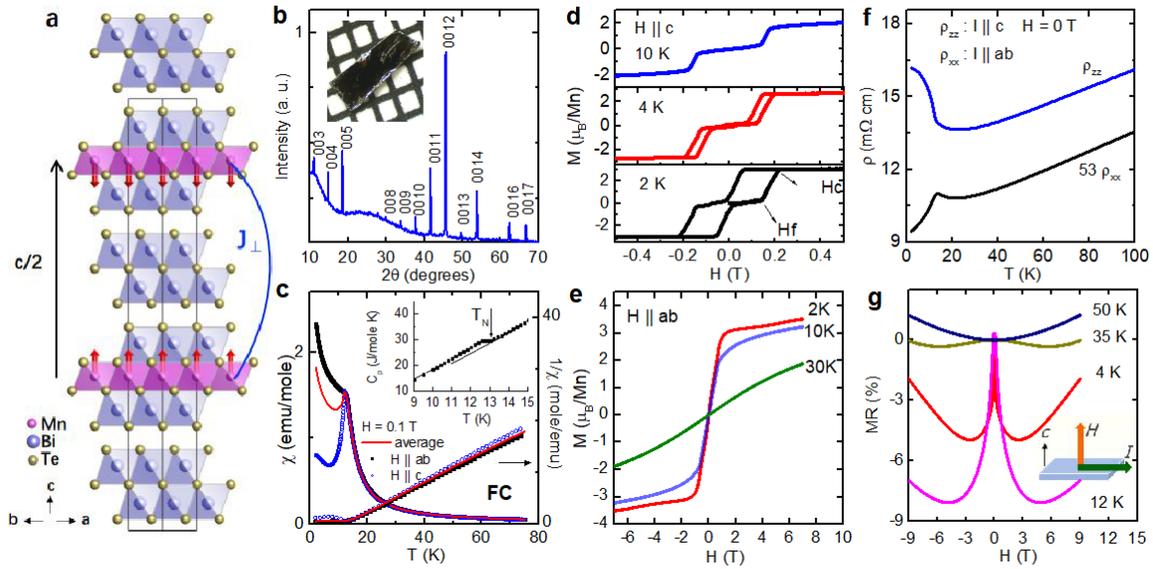

**Figure 1 | Magnetic and transport properties of bulk AFM MnBi$_4$Te$_7$.** **a** The view of the crystal structure of MnBi$_4$Te$_7$ from the [110] directions. Red arrow: Mn spins in the A-type AFM state. Blue block: edge-sharing BiTe$_6$ octahedra; Pink block: edge-sharing MnTe$_6$ octahedra, which are connected to the blue block via edge-sharing. J$_\perp$ is the interlayer exchange coupling. **b** The (00*l*) X-ray diffraction peaks of the cleaved *ab* plane of MnBi$_4$Te$_7$. Inset: A piece of MnBi$_4$Te$_7$ against 1-mm scale. **c** The temperature dependent field-cooled susceptibility and inverse susceptibility taken at *H* = 0.1 T for *H* ∥ *ab* and *H* ∥ *c*. Average $\chi$ is calculated by $\chi^{ave} = (2\chi^{ab} + \chi^{c})/3$. **d** and **e**: Full magnetic hysteresis loop of isothermal magnetization taken at various temperatures for: **d** *H* ∥ *c* and **e** *H* ∥ *ab*. **f** The temperature dependent $\rho_{xx}$ (*I* ∥ *ab*) and $\rho_{zz}$ (*I* ∥ *c*). **g** Transverse magnetoresistance with *I* ∥ *ab* and *H* ∥ *c* at various temperatures.

Figure 2.

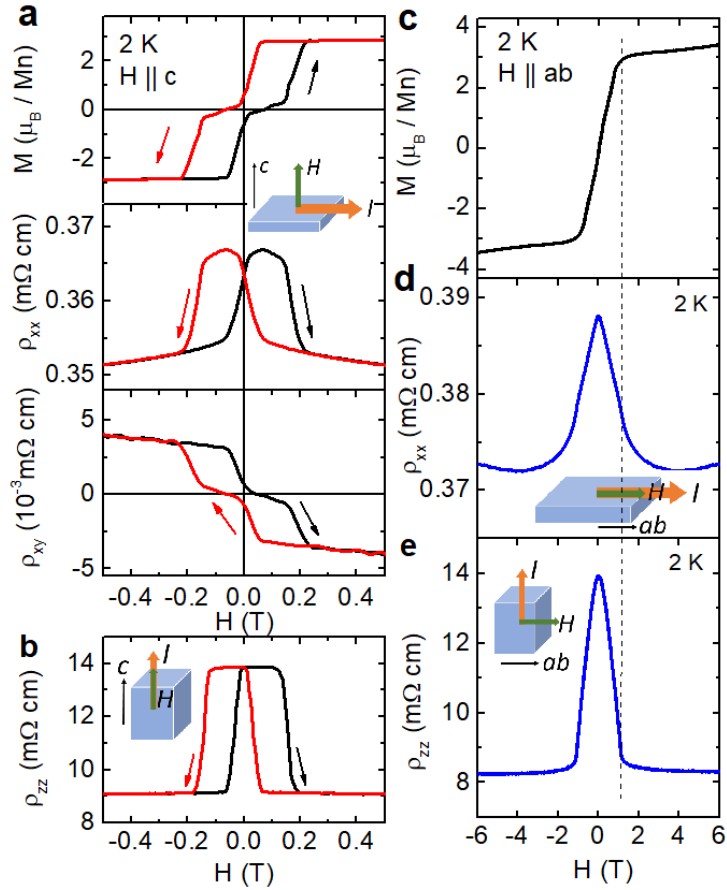

**Figure 2 | Magnetotransport of bulk AFM MnBi$_4$Te$_7$. a** The field dependent magnetization M, transverse magnetoresistivity $\rho_{xx}$, and Hall resistivity $\rho_{xy}$ at 2 K with $I \parallel ab$ and $H \parallel c$. **b** The longitudinal magnetoresistivity of $\rho_{zz}$, at 2 K with $I \parallel H \parallel c$. **c** The field dependent magnetization M with $H \parallel ab$ at 2 K. **d** The longitudinal magnetoresistivity $\rho_{xx}$, at 2 K with $I \parallel H \parallel ab$. **e** The transverse magnetoresistivity $\rho_{zz}$, at 2 K with $I \parallel c$ and $H \parallel ab$.

Figure 3.

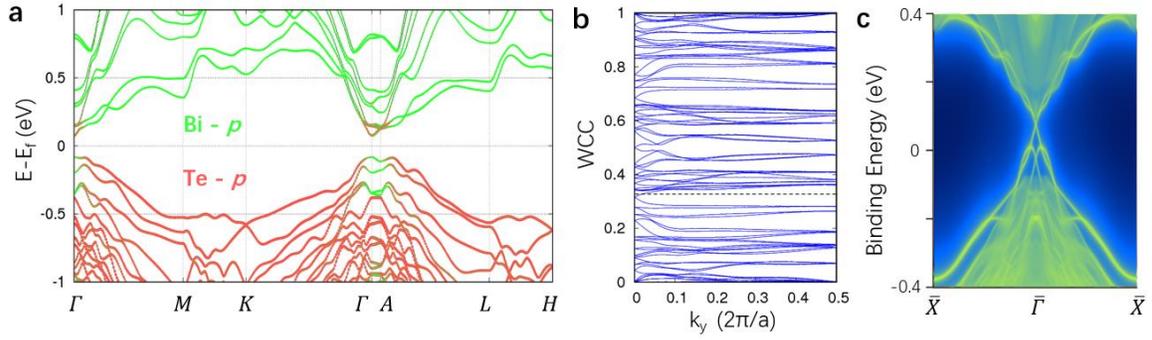

**Figure 3 | Topological properties of bulk AFM MnBi$_4$Te$_7$ predicted by first-principles calculations. a** Band structure with the projection of Bloch eigenstates onto Bi-*p* (green) and Se-*p* (red) orbitals. SOC is included. **b** Evolution of Wannier charge centers (WCCs) for $k_z = 0$, indicating a nontrivial topological invariant $Z_2 = 1$. **c** Surface spectra of (010) side surface, showing a gapless Dirac cone protected by *S* symmetry.

Figure 4.

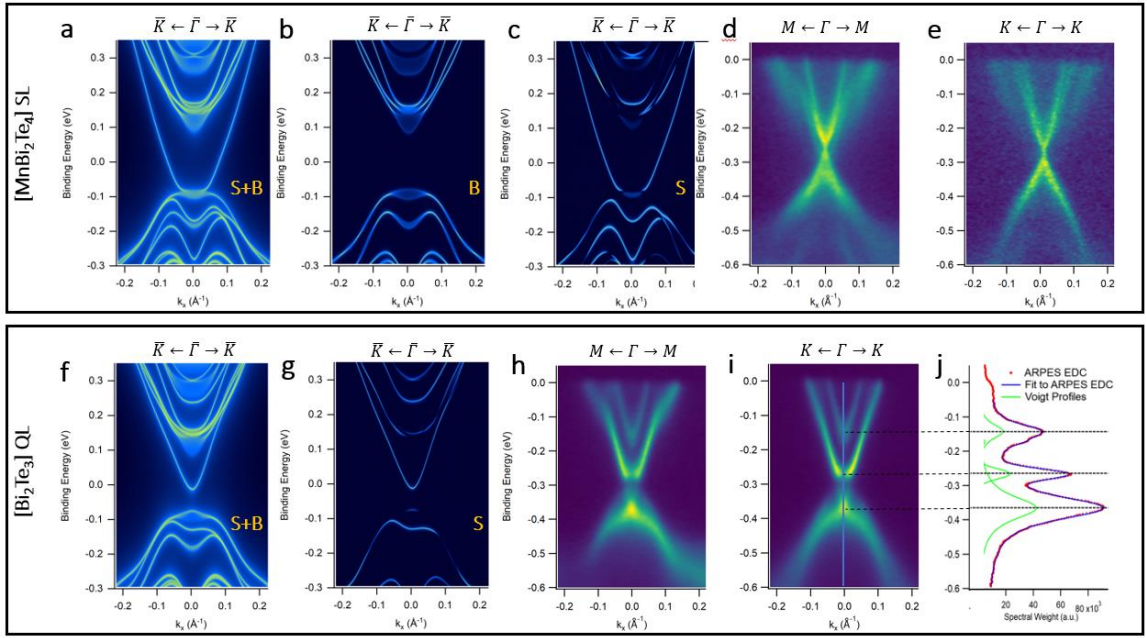

**Figure 4 | Comparison between ARPES-measured and calculated surface states. a-c** The DFT-calculated *k-E* map along $\bar{K} \leftarrow \bar{\Gamma} \rightarrow \bar{K}$ on the [MnBi$_2$Te$_4$] SL termination: **a** bulk and surface (B+S) spectrum, **b** bulk only, and **c** surface only. **d-e** The experimental ARPES spectrum on the [MnBi$_2$Te$_4$] SL termination obtained with 47 eV, linear horizontal light: **d** along $M \leftarrow \Gamma \rightarrow M$ **e** along $K \leftarrow \Gamma \rightarrow K$ high symmetry direction. **f-g** The DFT-calculated *k-E* map along $\bar{K} \leftarrow \bar{\Gamma} \rightarrow \bar{K}$ on the [Bi$_2$Te$_3$] QL termination: **f** bulk and surface spectrum, **g** surface only. **h-i** The experimental ARPES spectrum on the [Bi$_2$Te$_3$] QL termination obtained with 47 eV, linear horizontal light: **h** along $M \leftarrow \Gamma \rightarrow M$ **i** along $K \leftarrow \Gamma \rightarrow K$ high symmetry direction. **j** The EDC plot at the Γ point (blue-line cut in **i**) showing three main peaks corresponding to the bulk conduction band, surface conduction band, and mixed surface/bulk valence band. The green curve shows the fitted Voigt profile peaks which sum to the blue curve.

Figure 5.

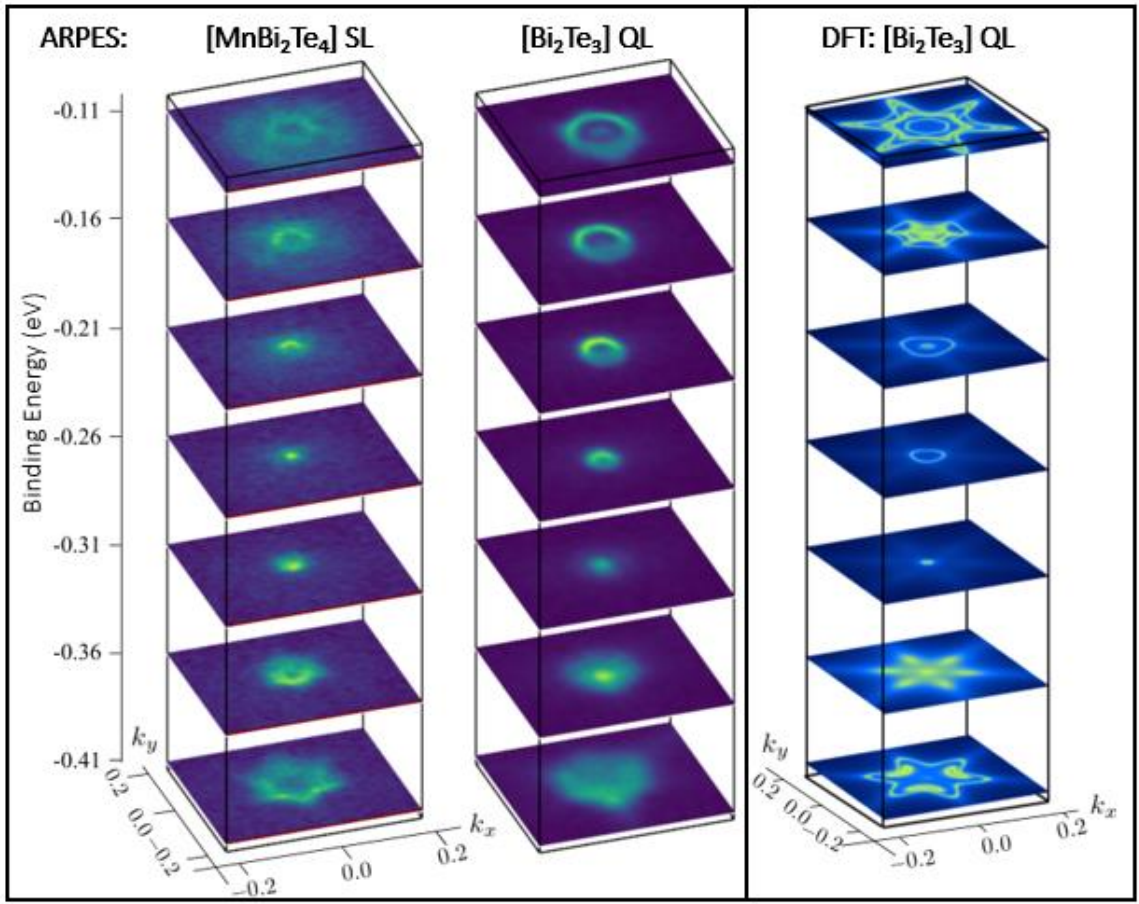

**Figure 5 | Experimental and theoretical constant energy slices.** (a,b) ARPES constant energy surfaces sliced at every 50 meV. (c) The same contours calculated by DFT for the [$Bi_2Te_3$] QL termination. The six-fold symmetric snowflake-like surfaces are seen in all cases.